\documentclass{JHEP3}
\usepackage{graphicx,amsmath,amssymb}
\usepackage{epsfig,multicol}
\usepackage{epsf}
\usepackage{epstopdf}
\input{epsf.sty}

\usepackage{graphicx}
\usepackage{amssymb}
\usepackage{amsmath}

\def\p{\partial}

\def\half{{1\over 2}}
\def\({\left(}
\def\){\right)}
\def\[{\left[}
\def\]{\right]}

\def\e{\begin{equation}}
\def\q{\end{equation}}
\def\m{\begin{eqnarray}}
\def\n{\end{eqnarray}}

\title{The trispectrum in ghost inflation}
\author{Qing-Guo Huang \footnote{huangqg@itp.ac.cn}
\\\small{\em
Kavli Institute for Theoretical Physics China (KITPC),
Key Laboratory of Frontiers in Theoretical Physics,
Institute of Theoretical Physics, Chinese Academy
of Sciences, Beijing 100190, China}
}

\abstract{
We calculate the trispectrum in ghost inflation where both the contact diagram and scale-exchange diagram are taken into account. The shape of trispectrum is discussed carefully and we find that the local form is absent in ghost inflation. In general, for the non-local shape trispectrum there are not analogous parameters to $\tau_{NL}^{loc.}$ and $g_{NL}^{loc.}$ which can completely characterize the size of local form trispectrum.
}

\keywords{inflation, non-gaussianity}

\begin{document}

\section{Introduction}

Even though inflation is an elegant paradigm in the early universe, a realistic inflation
model is still not known. A near scale-invariant primordial power spectrum can be taken as
one of the predictions of inflation. Nowadays the three-point and four-point correlation
functions of the curvature perturbations, so-called bispectrum and trispectrum, have been
a sensitive probe of the physics in the early universe. These correlation functions encode
rich information about the detail of inflation and may help us to understand the physics in
the early universe.

For simplicity, we classify the bispectrum and trispectrum to be
local and non-local shapes \cite{Babich:2004gb}. A large local-shape
non-Gaussianity can be generated at the end of inflation
\cite{Lyth:2005qk,Sasaki:2008uc,Huang:2009xa,Huang:2009vk} due to
the non-trivial condition for multi-field inflation to end, or deep
in the radiation era
\cite{Enqvist:2001zp,Lyth:2001nq,Moroi:2001ct,Sasaki:2006kq,Huang:2008ze,Huang:2008rj,Huang:2008zj}.
See other relevant papers in \cite{refs}. In general the multi
fields must be involved in order to generate large
local-shape bispectrum and trispectrum. The local-shape non-Gaussianity is much more
sensitive to the experiments.
WMAP 7yr data \cite{Komatsu:2010fb} implies $-10<f_{NL}^{loc.}<74$
at $95\%$ C.L.. A convincing detection of such a large local-shape
non-Gaussianity will rule out all single-field inflation models.

However a large non-local form non-Gaussianity is also allowed by
WMAP 7yr data \cite{Komatsu:2010fb,Huang:2010up}. The mechanism for
producing a non-local form non-Gaussianity is quite different from a
local form non-Gaussianity. The non-local form non-Gaussiniaty is
generated on the horizon scale during inflation, and all of the
relevant perturbation modes have roughly the same wavelength. In
general, higher derivative terms in the
action of the inflaton field are called for in order to produce a large non-local form bispectrum and trispectrum. The mixed scenario in
\cite{Huang:2010up} opens a window to achieve not only local form
but non-local form non-Gaussianty in one model where the higher
derivative terms for the inflaton along the adiabatic direction
contributes to a large non-local shape non-Gaussianity, and the
other light scalar fields, such as curvaton(s), produce a large
local shape non-Gaussianity. A distinguishing phenomenology for the
mixed scenario is that $\tau_{NL}^{loc.}$ must be larger than
$({6\over 5}f_{NL}^{loc.})^2$.


In this paper we focus on the primordial trispectrum in the ghost inflation model \cite{ArkaniHamed:2003uz} based on the idea of ghost condensate \cite{ArkaniHamed:2003uy}. We will see that the local shape trispectrum is absent in this single field model. The ghost condensate is a new kind of fluid that can fill the universe. It has the equation of state $p=-\rho$ and provides an alternative way of realizing de Sitter phases in the early universe. The ghost scalar field $\phi$ condenses in a background where it has a non-zero velocity
\e
\langle \dot \phi \rangle=M^2,
\q
where $M$ is the ghost cur-off scale and the scalar field $\phi$ has a constant velocity $\langle \dot \phi \rangle$ from ghost condensation. Several observational constraints on the cut-off scale, for example $M<100$ GeV, are discussed in \cite{ArkaniHamed:2003uz,ArkaniHamed:2003uy}.
The ghost condensate is not a cosmological constant, it is a physical fluid with a physical scalar excitation $Q$ defined as
\e
\phi=M^2 t+Q.
\q
Assuming that the ghost field $\phi$ has a shift symmetry $\phi\rightarrow \phi+\delta$, we conclude that the ghost field must be derivatively coupled. Based on the symmetry of this theory, we construct the action for $Q$ as follows
\m
S=\int d^4x \sqrt{-g} \[ \half {\dot Q}^2 -{\alpha^2\over 2M^2}(\nabla^2 Q)^2 +{\cal L}_{int}^{(3)} + {\cal L}_{int}^{(4)}+... \] \ ,
\label{acs}
\n
where
\m
{\cal L}_{int}^{(3)}&=&-{\beta_1\over 2M^2}\dot Q (\nabla Q)^2-{\beta_2\over 2M^3}\nabla^2 Q (\nabla Q)^2+...\ , \\
{\cal L}_{int}^{(4)}&=&-{{\tilde \gamma} \over 8M^4}(\nabla Q)^4+...\ .
\n
Here we only take the leading order interaction terms into account. We need to remind that the term of $\nabla^2 Q (\nabla Q)^2$ breaks the symmetry $Q\rightarrow -Q$ and $t\rightarrow -t$ which corresponds to $\phi\rightarrow -\phi$ for the ghost Lagrangian, and one may worry about the violation of CPT symmetry. However, the size of the CPT violating operator depends on the coupling of the ghost sector to ordinary matter and it is not predicted. The bispectrum from this term is expected to be similar to that from $\dot Q (\nabla Q)^2$ \cite{ArkaniHamed:2003uz}. In this paper, we keep this term in the action and calculate the trispectrum related to this term as well.

Our results are not included in the previous discussions about the
trispectrum in the general Lorentz invariant single-field inflation
\cite{Seery:2006vu,Seery:2006js,Arroja:2008ga,Seery:2008ax,Chen:2009bc,Arroja:2009pd,Cogollo:2008bi}
and multi-field cases
\cite{Gao:2009gd,Mizuno:2009cv,Gao:2009at,Mizuno:2009mv}, because
the action for the quantum fluctuation $Q$ breaks the Lorentz
invariance. Our results are reliable at energies lower than the
ghost cur-off $M$.

This paper is organized as follows. In Sec. 2 we repeat the calculation of the primordial power spectrum which is necessary for us to work out the non-Gaussianity parameters. In Sec. 3 we compute the trispectrum contributed from the contact and scalar-exchange diagram in ghost inflation. The conclusion and some discussions are contained in Sec. 4.

\section{The Hamiltonian density in the interaction picture}

From (\ref{acs}), the conjugate field $\Pi$ in Heisenberg picture is defined as
\e
\Pi={\p {\cal L}\over \p \dot Q}=\dot Q-{\beta_1\over 2M^2}(\nabla Q)^2,
\q
and then $\dot Q$ is related to the conjugate field by
\e
\dot Q=\Pi+{\beta_1\over 2M^2}(\nabla Q)^2.
\q
The Hamiltonian density becomes
\m
{\cal H}&=&\dot Q \Pi-{\cal L} \nonumber \\
&=&\half \Pi^2+{\alpha^2\over 2M^2}(\nabla^2 Q)^2+{\beta_1\over 2M^2}\Pi (\nabla Q)^2+{\beta_2\over 2M^3}\nabla^2Q (\nabla Q)^2+{\gamma\over 8M^4}(\nabla Q)^4,
\n
where
\e
\gamma={\tilde \gamma}+\beta_1^2.
\q
In order to calculate four-point correlation function, we switch to the interaction representation and quantize this field theory.
The fields $Q$ and $\Pi$ should be replaced by $Q_I$ and $\Pi_I$ in the interaction picture.
Here the subscript `$I$' denotes that the operators are in the
interaction representation.
The free-particle Hamiltonian density is adopted as
\e
{\cal H}_0=\half \Pi_I^2+{\alpha^2\over 2M^2}(\nabla Q_I)^2,
\q
and the commutation relation between operators $Q_I$ and
$\Pi_I$ is given by
\e
[Q_I, \Pi_I]=i.
\q
The time derivative of $Q_I$ is
\e
\dot Q_I=-i[Q_I, {\cal H}_0]={\p {\cal H}_0\over \p \Pi_I}=\Pi_I.
\q
Therefore the interaction Hamiltonian density becomes
\m
{\cal H}_I={\cal H}-{\cal H}_0={\beta_1\over 2M^2}{\dot Q}_I (\nabla Q_I)^2+{\beta_2\over 2M^3}\nabla^2Q_I (\nabla Q_I)^2+{\gamma\over 8M^4}(\nabla Q_I)^4.
\n
From now on we omit the subscript `$I$' in the variables.
\footnote{Our result is different from that in \cite{Izumi:2010wm} where $\gamma={\tilde \gamma}+2\beta_1^2$.}

\section{The power spectrum}

In \cite{ArkaniHamed:2003uy}, the authors showed that the gravitational potential $\Phi$ decays to zero outside the horizon. It implies that the fluctuations of $Q$ do not gravitate at the superhorizon scales in the pure de Sitter space. Here we focus on the gauge invariant quantity which is related to $Q$ by
\e
\zeta=-{H\over \dot \phi}Q=-{H\over M^2}Q.
\q
This quantity is conserved outside the horizon and it will seed the temperature fluctuations in the cosmic microwave background radiation.

In presence of ghost condensate gravity is modified in the IR region. This modification is characterized by a typical time scale $\Gamma^{-1}$ with $\Gamma\sim M^3/M_p^2$, and a typical length scale $m^{-1}$ with $m\sim M^2/M_p$ \cite{ArkaniHamed:2003uy}. Requiring that gravity is not modified nowadays on the scales smaller than the present Hubble horizon, we need to impose $\Gamma<H_0$, where $H_0$ is present Hubble parameter. Therefore we have $\Gamma\ll m\ll H$, and the gravity is not modified during inflation.

From the action (\ref{acs}), at the linear level, we have
\e
\ddot Q+3H\dot Q+{\alpha^2\over M^2}\nabla^4 Q=0,
\q
where
\e
\nabla^2=g^{ij}\p_i\p_j={1\over a^2}\p_i\p_i.
\q
The field $Q$ is canonically quantized as
\e
Q({\bf x},t)=\int {d^3k\over (2\pi)^3}Q_{\bf k}(t)e^{i{\bf k}\cdot {\bf x}},
\q
with
\e
Q_{\bf k}(t)=q_k(t){\hat {a}}_{\bf k} +q_k^*(t){\hat {a}}_{-\bf k}^+,
\q
and
\e
[{\hat {a}}_{\bf k}, {\hat a}_{-{\bf k}'}^+]=(2\pi)^3 \delta^{(3)}({\bf k}+{\bf k}').
\q
Here ${\bf k}$ is the comoving wavevector, and $q_k(t)$ is governed by
\e
\ddot q_k+3H\dot q_k+{\alpha^2\over M^2}{k^4\over a^4}q_k=0.
\q
We introduce a new time coordinate, so-called conformal time $\tau$ which is related to $t$ by
\e
d\tau={dt\over a(t)}.
\q
For an inflationary universe, we have
\e
a=-{1\over H\tau},
\q
where $H$ is the Hubble parameter which can be taken as a constant in ghost inflation. Now the equation for the variable
\e
u_k=a(\tau)\cdot q_k
\q
becomes
\e
u_k''+\({\alpha^2H^2k^4\over M^2}\tau^2-{2\over \tau^2}\)u_k=0,
\q
where the prime denotes the derivative with respect to the conformal time $\tau$. The solution of the above differential equation with the correct flat space limit for the very short physical wave-length $(\tau\rightarrow -\infty)$ is given by
\e
u_k(\tau)=\sqrt{\pi\over 8}\sqrt{-\tau}H_{3/4}^{(1)}\({\alpha Hk^2\over 2M}\tau^2\),
\q
where $H_\nu^{(1)}$ is the Hankel function of the first kind. Considering $q_k=u_k/a(\tau)=-H\tau\cdot u_k$, we obtain
\e
q_k(\tau)=\sqrt{\pi\over 8}H\cdot (-\tau)^{3/2} H_{3/4}^{(1)}\({\alpha Hk^2\over 2M}\tau^2\).
\q

The power spectrum can be calculated from the asymptotic behavior in the limit of $\tau\rightarrow 0$. In this limit, we find
\e
q_k(\tau\rightarrow 0)\simeq -{i\sqrt{2\pi}\over \Gamma(1/4)} H \({M\over \alpha H}\)^{3/4}k^{-3/2}.
\q
From the definition of power spectrum $P_\zeta(k)$, namely
\e
\langle \zeta_{{\bf k}} \zeta_{{\bf k}'} \rangle=(2\pi)^3\delta^{(3)} ({\bf k}+{\bf k}'){2\pi^2\over k^3}P_\zeta(k),
\q
we can easily calculate the amplitude of the primordial curvature perturbation,
\e
P_\zeta={1\over \pi \Gamma^2(1/4)}\({H\over M}\)^{5/2}\alpha^{-3/2}.
\q
Our results is the same as that in \cite{ArkaniHamed:2003uz}.
Using the WMAP normalization $P_\zeta=2.41\times 10^{-9}$ \cite{Komatsu:2010fb}, we obtain
\e
{H\over M}\simeq 1.58\times 10^{-3}\alpha^{3/5}.
\label{wmapn}
\q
Since $\Gamma\sim M^3/M_p^2\ll H_0$ which implies $M\lesssim 10^{-20}M_p$, the energy scale of inflation is so small compared to Planck scale that the gravitational wave perturbation in ghost inflation must be completely negligible.

\section{The trispectrum}

In this section, we will explicitly calculate the four-point correlation function of the curvature perturbation in ghost inflation and discuss the shape of trispectrum.

A well established non-Gaussianity has a local shape.
For the local shape non-Gaussianity, the curvature perturbation $\zeta$ can be expanded to the non-linear orders at the same spatial point,
\m
\zeta({\bf x})=\zeta_g({\bf x})+{3\over 5}f_{NL}^{loc.}\zeta_g^2({\bf x})+{9\over 25} g_{NL}^{loc.} \zeta_g^3({\bf x})+... \ ,
\n
where $\zeta_g$ is the Gaussian part of curvature perturbation. The four-point irreducible correlation function of curvature perturbations are related to the power spectrum by
\m
\langle \zeta_{{\bf k}_1} \zeta_{{\bf k}_2} \zeta_{{\bf k}_3} \zeta_{{\bf k}_4} \rangle=(2\pi)^9 \delta^{(3)}(\sum_{i=1}^4{\bf k}_i) T_\zeta(k_1,k_,k_3,k_4),
\n
where
\m
T_\zeta(k_1,k_2,k_3,k_4)&=& {27\over 100}g_{NL}^{loc.}P_\zeta^3 \cdot {\sum_{i=1}^4 k_i^3\over \prod_{i=1}^4 k_i^3} \\
&+&  {1\over 16} \tau_{NL}^{loc.} P_\zeta^3 \cdot \({1 \over k_{12}^3 k_2^3 k_3^3}+23 \ \hbox{perms}.\),
\label{tzeta}
\n
and $k_{12}=k_{34}$, $k_{13}=k_{24}$ etc. Here
\e
{\bf k}_{ij}={\bf k}_i+{\bf k}_j.
\q
If $g_{NL}^{loc.}\neq 0$, the term of $g_{NL}^{loc.}$ blows up when one or two of $k_i$ goes to zero.
If the curvature perturbation is generated by single scalar field, we get a consistency relation, namely
$\tau_{NL}^{loc.}=({6\over 5}f_{NL}^{loc.})^2$.
If $\tau_{NL}^{loc.}>({6\over 5}f_{NL}^{loc.})^2$, the curvature perturbation must be generated by more than one scalar fields. From (\ref{tzeta}), the term with $\tau_{NL}^{loc.}$ also blows up when one or two of $k_i$ goes to zero. The distinguishing feature for the term of $\tau_{NL}^{loc.}$ is that the local form trispectrum blows up in the limit of $k_{ij}\rightarrow 0$, even when $k_1=k_2=k_3=k_4\neq 0$. Because the local shape trispectrum does blow up for these special configurations in the momenta space, it is much more sensitive to the cosmological observations than that with non-local shape.

\begin{figure}[h]
\begin{center}
\includegraphics[width=5cm]{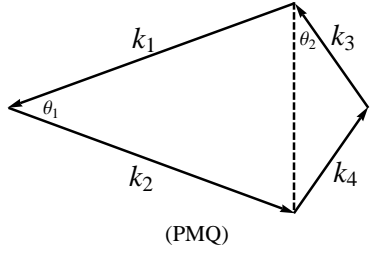}
\end{center}
\caption{The planar mirror symmetric quadrangle shape. }
\label{pmq}
\end{figure}
\begin{figure}[h]
\begin{center}
\includegraphics[width=5cm]{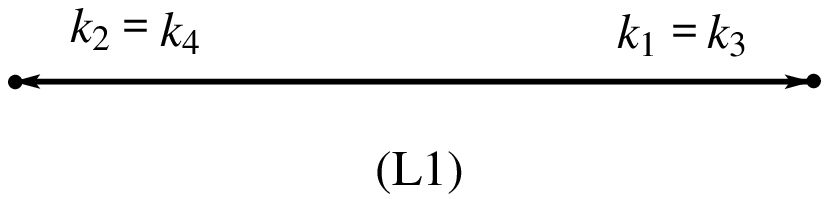}\\
\vspace{.5cm}
\includegraphics[width=5cm]{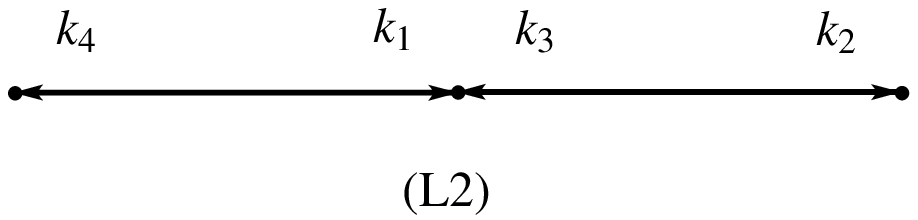}\\
\vspace{.5cm}
\includegraphics[width=5cm]{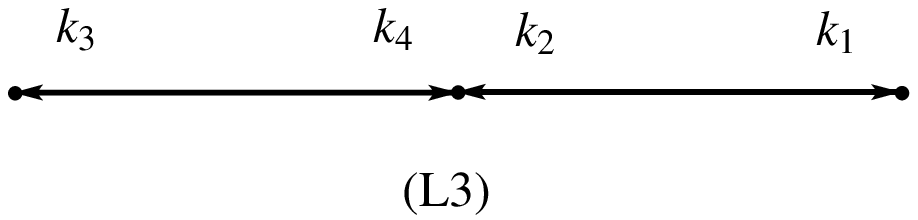}\\
\vspace{.5cm}
\includegraphics[width=5cm]{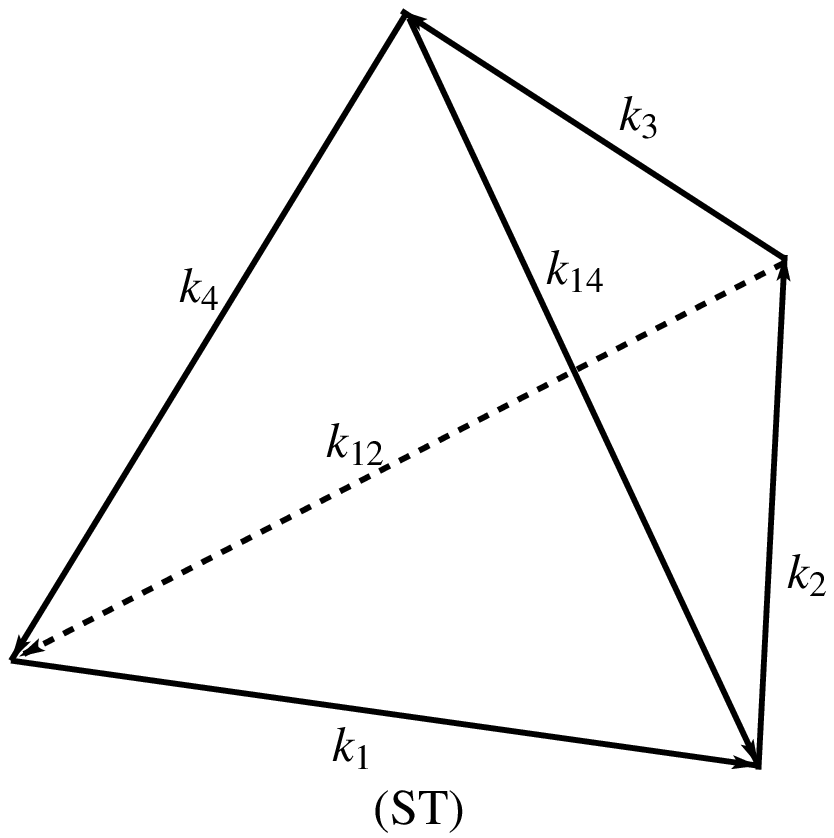}
\end{center}
\caption{The special configurations in the momenta space for the trispectrum. }
\label{spt}
\end{figure}

The energy-momentum conservation implies that the four momenta vectors ${\bf k}_i$ ($i=1,2,3,4$) form a quadrangle which is much more complicated than the triangle for the bispectrum. Its shape cannot be fixed even when the sizes of these four vectors are fixed. In this paper, we suggest a few special configurations. \\
\noindent $\bullet$ Planar mirror symmetric quadrangle. See Fig. \ref{pmq}.
Here $\theta_1\in[0,\pi]$ and $\theta_2\in[0,\pi/2]$,
\e
k_1=k_2=k,\quad k_3=k_4={\sin {\theta_1\over 2}\over \cos \theta_2}k
\q
and
\m
{{\bf k}_1\cdot {\bf k}_2 \over k^2}&=&-\cos\theta_1, \quad {{\bf k}_3\cdot {\bf k}_4 \over k^2}=\({\sin {\theta_1\over 2}\over \cos \theta_2}\)^2 \cos(2\theta_2), \\
{{\bf k}_1\cdot {\bf k}_3 \over k^2}&=&{{\bf k}_2\cdot {\bf k}_4 \over k^2}=-{\sin {\theta_1\over 2}\over \cos \theta_2}\sin({\theta_1\over 2}+\theta_2), \\
{{\bf k}_1\cdot {\bf k}_4 \over k^2}&=&{{\bf k}_2\cdot {\bf k}_3 \over k^2}=-{\sin {\theta_1\over 2}\over \cos \theta_2}\sin({\theta_1\over 2}-\theta_2).
\n
If $\theta_2=\pi/3$, in the limit of $\theta_1\rightarrow 0$, $k_3=k_4=k_{12}=k_{34}\simeq \theta_1 k$ and the local form trispectrum becomes
\m
T_\zeta(k_1,k_2,k_3,k_4)\simeq \half \({27\over 25}g_{NL}^{loc.}+\tau_{NL}^{loc.}\){P_\zeta^3\over \theta_1^6k^9}
\n
which goes to infinity. If $\theta_2=(\pi-\theta_1)/2$, $k_1=k_2=k_3=k_4$. In this case $k_{12}=k_{34}\simeq \theta_1 k$ in the limit of $\theta_1\rightarrow 0$, and hence
\e
T_\zeta(k_1,k_2,k_3,k_4)\simeq \half \tau_{NL}^{loc.}{P_\zeta^3\over \theta_1^3k^9}
\q
which also blows up for non-vanishing $\tau_{NL}^{loc.}$. \\
\noindent $\bullet$ Equilateral shape. Now $k_1=k_2=k_3=k_4\equiv k$, but there are still two extra degrees of freedom: the angle $\theta_{12}$ between the vector ${\bf k}_1$ and ${\bf k}_2$, and the angle $\theta_{14}$ between the vector ${\bf k}_1$ and ${\bf k}_4$. The angle between ${\bf k}_1$ and ${\bf k}_3$ is not an independent parameter which is related to $\theta_{12}$ and $\theta_{14}$ by
\e
\theta_{13}=\pi-\cos^{-1}(1+\cos\theta_{12}+\cos\theta_{14}).
\q
In this case we have
\m
{{\bf k}_1\cdot {\bf k}_2 \over k^2}&=&{{\bf k}_3\cdot {\bf k}_4\over k^2}=\cos\theta_{12},\\
{{\bf k}_1\cdot {\bf k}_3 \over k^2}&=&{{\bf k}_2\cdot {\bf k}_4\over k^2}=\cos\theta_{13},\\
{{\bf k}_1\cdot {\bf k}_4 \over k^2}&=&{{\bf k}_2\cdot {\bf k}_3\over k^2}=\cos\theta_{14}.
\n
The consistency condition for formation of a quadrangle is given by
\e
\cos\theta_{12}+\cos\theta_{14}\leq 0.
\q
Because all of the momenta vectors are finite, the term of $g_{NL}^{loc.}$ in the local form trispectrum is still finite, but the term of $\tau_{NL}^{loc.}$ can go to infinity when one or two of $\theta_{1i}$ ($i=2,3,4$) goes to $\pi$. For example, if $\theta_{12}\rightarrow \pi$, we find
\e
T_\zeta(k_1,k_2,k_3,k_4)\simeq \half \tau_{NL}^{loc.}{P_\zeta^3\over (\pi-\theta_{12})^3k^9}.
\q
Here we suggest several special equilateral configurations, named L1, L2, L3, and ST (special tetrahedron) in Fig. \ref{spt}.
The angles $\theta_{12}$ and $\theta_{14}$ take the values for these configurations as follows
\m
\hbox{L1}:&& \quad \theta_{12}=\theta_{14}=\pi,\\
\hbox{L2}:&& \quad \theta_{12}=0,\ \theta_{14}=\pi,\\
\hbox{L3}:&& \quad \theta_{12}=\pi,\ \theta_{14}=0,\\
\hbox{ST}:&& \quad \theta_{12}=\theta_{14}=\cos^{-1}(-1/3).
\n
For the shape ``ST", $k_{12}=k_{14}=k_{13}=2k/\sqrt{3}$.

At the leading order, not only ${\cal H}_{int}^{(4)}$ but also ${\cal H}_{int}^{(3)}$ contribute to the trispectrum in ghost inflation. We will calculate the trispectrum generated by these two interaction terms in the following two subsections respectively.

\subsection{Contact diagram}

\begin{figure}[h]
\begin{center}
\includegraphics[width=5cm]{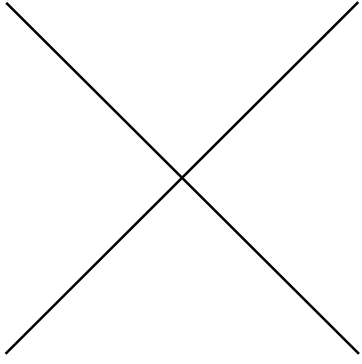}
\end{center}
\caption{The contact interaction. }
\label{contact}
\end{figure}

In this subsection, we compute the four-point correlation function from the contact interaction diagram illustrated in Fig. \ref{contact}.
This diagram corresponds to the interaction term ${\cal L}_{int}^{(4)}$.
In this case, $\langle Q_{{\bf k}_1} Q_{{\bf k}_2} Q_{{\bf k}_3} Q_{{\bf k}_4} \rangle$ is given by
\m
\langle Q_{{\bf k}_1}(t) Q_{{\bf k}_2}(t) Q_{{\bf k}_3}(t) Q_{{\bf k}_4}(t) \rangle \supset -i\int_{t_0}^t dt' \left\langle \[Q_{{\bf k}_1}(t) Q_{{\bf k}_2}(t) Q_{{\bf k}_3}(t) Q_{{\bf k}_4}(t), H_I^{(4)} (t')\] \right\rangle, \nonumber \\
\n
where
\e
H_I^{(4)}=\int d^3 x {\cal H}_{int}^{(4)},
\q
and
\e
{\cal H}_{int}^{(4)}={\gamma \over 8M^4}{1\over a(t)}(\p_i Q\cdot \p_i Q)^2.
\q
Considering
\e
\zeta_{\bf k}=-{H\over M^2}Q_{\bf k},
\q
we obtain
\m
\langle \zeta_{{\bf k}_1} \zeta_{{\bf k}_2} \zeta_{{\bf k}_3} \zeta_{{\bf k}_4} \rangle &\supset& -i (2\pi)^3\delta^{(3)}(\sum_{i=1}^4 {\bf k}_i) \cdot \gamma \cdot {H^4\over M^{12}} \cdot  \prod_{i=1}^{4} q_{k_i}(0) \nonumber \\
&\times& \int_{-\infty}^0 d\tau \cdot \prod_i q_{k_i}^*(\tau) \cdot ({\bf k}_1\cdot {\bf k}_2) ({\bf k}_3\cdot {\bf k}_4) \nonumber \\
&+&\hbox{symm}.+c.c.
\n
which is evaluated as
\m
\langle \zeta_{{\bf k}_1} \zeta_{{\bf k}_2} \zeta_{{\bf k}_3} \zeta_{{\bf k}_4} \rangle &\supset& + i (2\pi)^9\delta^{(3)}(\sum_{i=1}^4 {\bf k}_i) {\gamma\over 2^{10} \cdot \pi^2 \cdot \Gamma^4(1/4)} \({H\over M}\)^{11\over 2} \alpha^{-{13\over 2}} {({\bf k}_1\cdot {\bf k}_2) ({\bf k}_3\cdot {\bf k}_4) \over k^7 \prod_{i}k_i^{3/2}} \nonumber \\
&\times& \int_{-\infty}^0 dx\cdot x^6 \cdot H_{3/4}^{(1)}({k_1^2\over k^2}{x^2\over 2})  H_{3/4}^{(1)}({k_2^2\over k^2}{x^2\over 2})  H_{3/4}^{(1)}({k_3^2\over k^2}{x^2\over 2}) H_{3/4}^{(1)}({k_4^2\over k^2}{x^2\over 2}) \nonumber \\
&+&\hbox{symm}.+c.c. \ ,
\n
where
\e
x=\({\alpha H\over M}\)^\half k\tau.
\q
This is the full trispectrum from the contact interaction diagram. It is very difficult to get the analytic result. We will evaluate the numerical results for the special shapes in our previous discussions.
Comparing to the local form trispectrum, we effectively define the non-Gaussianity parameters $g_{NL}$ as follows
\m
T_\zeta(k_1,k_2,k_3,k_4) \supset {27\over 100}g_{NL} P_\zeta^3 \cdot {\sum_{i=1}^4 k_i^3\over \prod_{i=1}^4 k_i^3}.
\label{defgnl}
\n
In general $g_{NL}$ is not a constant, but a function which depends on $k_i$. For the non-local shape trispectrum, $g_{NL}$ goes to zero in the limit where one or two of $k_i$ go to zero.

\begin{figure}[h]
\begin{center}
\includegraphics[width=10cm]{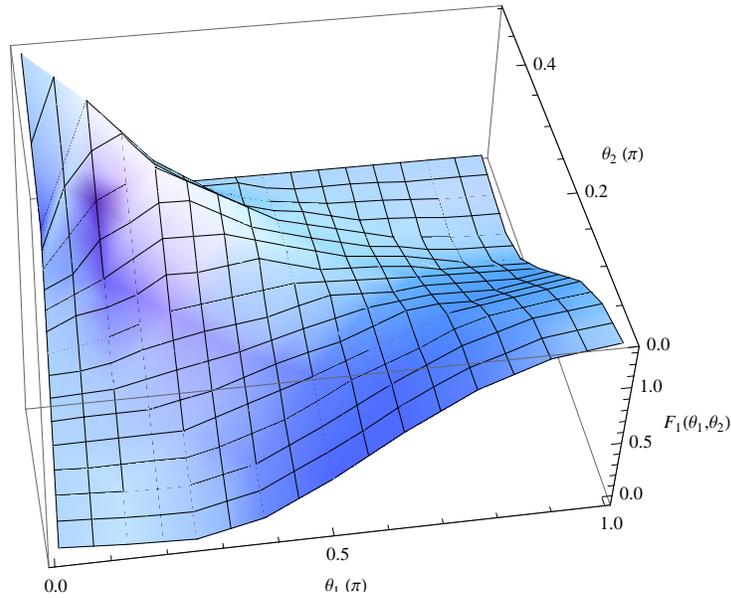}
\end{center}
\caption{The function $F_1(\theta_1,\theta_2)$. }
\label{gtpmq}
\end{figure}

For the planar mirror symmetric quadrangle shape in Fig. \ref{pmq}, the ``effective" non-Gaussianity parameter $g_{NL}^{pmq}$ depends on the angles $\theta_1$ and $\theta_2$,
\m
g_{NL}^{pmq}=-0.149  \cdot \gamma\cdot {M^2\over H^2}\alpha^{-2} \cdot F_1(\theta_1,\theta_2),
\n
where $F_1(\theta_1,\theta_2)$ shows up in Fig. \ref{gtpmq}.
Combing with the WMAP normalization (\ref{wmapn}), we obtain
\e
g_{NL}^{pmq}=-6\times 10^4  \cdot \gamma\cdot \alpha^{-16/5} \cdot F_1(\theta_1,\theta_2).
\q
For $\theta_2=\pi/3$, $g_{NL}^{pmq}\rightarrow 0$ in the limit of $\theta_1\rightarrow 0$. The local form trispectrum cannot be generated by the contact diagram in ghost inflation.

\begin{figure}[h]
\begin{center}
\includegraphics[width=10cm]{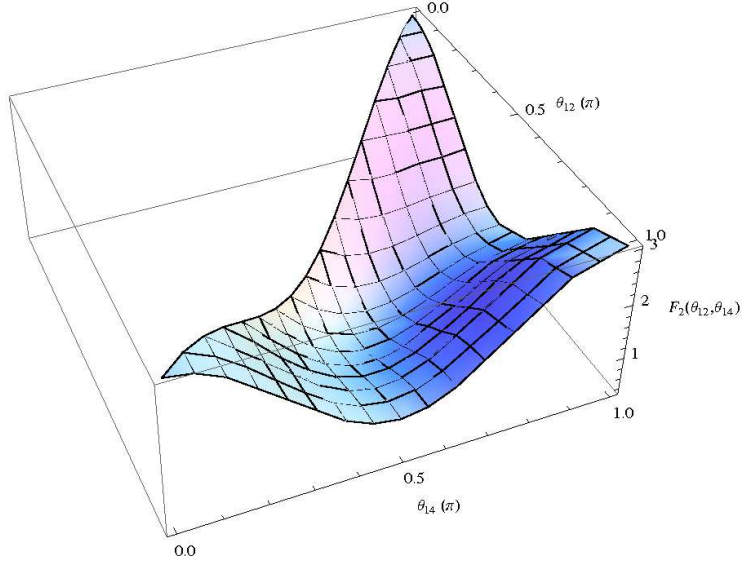}
\end{center}
\caption{The function of $F_2(\theta_{12},\theta_{14})$. }
\label{ff}
\end{figure}

Similarly, the ``effective" $g_{NL}^{equil.}$ for the equilateral shape is given by
\m
g_{NL}^{equil.}\simeq -0.066  \cdot \gamma\cdot {M^2\over H^2}\alpha^{-2} \cdot F_2(\theta_{12},\theta_{14}),
\n
where
\e
F_2(\theta_{12},\theta_{14})=\cos^2\theta_{12}+\cos^2\theta_{14}+(1+\cos\theta_{12}+\cos\theta_{14})^2.
\q
The angle-dependent function $F(\theta_{12},\theta_{14})$ is illustrated in Fig. \ref{ff}.
Combing with the WMAP normalization (\ref{wmapn}), we obtain
\m
g_{NL}^{equil.}
\simeq -2.6\times 10^4  \cdot \gamma\cdot \alpha^{-16/5} \cdot F_2(\theta_{12},\theta_{14}),
\n
For the special shapes in Fig. \ref{spt}, we have
\m
F_2^{L1}&=&F_2^{L2}=F_2^{L3}=3, \\
F_2^{ST}&=&1/3.
\n
We see that the size of equilateral shape trispectrum from the contact interaction is maximized when these four equilateral momenta vectors lie in a straight line and minimized at the ``ST" shape.

\subsection{Scalar-exchange diagram}

\begin{figure}[h]
\begin{center}
\includegraphics[width=5cm]{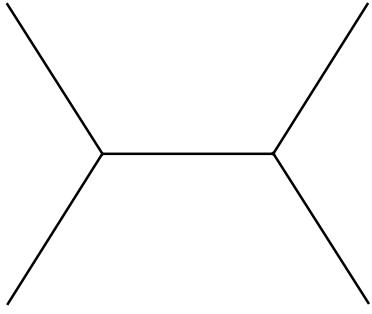}
\end{center}
\caption{The scalar-exchange diagram. }
\label{exc}
\end{figure}

In this subsection, we switch to a more complicated case: the scalar-exchange diagram in Fig. \ref{exc}.
There are two vertices in this diagram. Each of them can come from the term characterized by $\beta_1$ or $\beta_2$ in ${\cal L}_{int}^{(3)}$.
The four-point correlation function in the scalar-exchange diagram is given by, \cite{Adshead:2009cb},
\m
\langle Q_{{\bf k}_1}  Q_{{\bf k}_2}  Q_{{\bf k}_3}  Q_{{\bf k}_4} \rangle &\supset& \int_{t_0}^t dt_1 \int_{t_0}^t dt_2 \langle 0| H_I^{(3)}(t_1) Q_{{\bf k}_1} Q_{{\bf k}_2} Q_{{\bf k}_3} Q_{{\bf k}_4} H_I^{(3)}(t_2)|0\rangle \label{hqh} \\
&-&2\hbox{Re} \ \int_{t_0}^t dt_1 \int_{t_0}^{t_1} dt_2 \langle 0| H_I^{(3)}(t_2) H_I^{(3)}(t_1) Q_{{\bf k}_1}  Q_{{\bf k}_2} Q_{{\bf k}_3} Q_{{\bf k}_4} |0\rangle,   \label{hhq}
\n
where
\e
H_I^{(3)}=\int d^3x {\cal H}_{int}^{(3)},
\q
and
\m
{\cal H}_{int}^{(3)}=-{\cal L}_{int}^{(3)}={\beta_1\over 2M^2} {d\over d\tau}Q \cdot \p_i Q\cdot \p_i Q+{\beta_2\over 2M^3}{1\over a(t)}\p_i \p_i Q\cdot \p_j Q\cdot \p_j Q.
\n
For simplicity, we assume that both of the vertices are contributed by the interaction term with $\beta_1$ firstly. We write down the results of Eq.(\ref{hqh}) and (\ref{hhq}) separately.
Eq.(\ref{hqh}) is
\m
&&(2\pi)^3\delta^{(3)}(\sum_{i=1}^4{\bf k}_i) \cdot {\beta_1^2\over 4M^4H^2} \cdot q_{k_1}^*(0) q_{k_2}^*(0) q_{k_3}(0)q_{k_4}(0) \times \nonumber \\
&&\left\{ \int_{-\infty}^0 {d\tau_1\over \tau_1} {d\over d\tau_1} q_{k_{12}}(\tau_1)\cdot q_{k_1}(\tau_1)q_{k_2}(\tau_1) \right. \nonumber \\
&&\times \left[ ({\bf k}_1\cdot {\bf k}_2) ({\bf k}_3\cdot {\bf k}_4)  \int_{-\infty}^0 {d\tau_2\over \tau_2} \cdot {d\over d\tau_2} q_{k_{12}}^* (\tau_2)\cdot q_{k_3}^*(\tau_2)q_{k_4}^*(\tau_2)  \right. \nonumber \\
&&\left. + 2 ({\bf k}_1\cdot {\bf k}_2) ({\bf k}_{12}\cdot {\bf k}_4)  \int_{-\infty}^0 {d\tau_2\over \tau_2} \cdot q_{k_{12}}^* (\tau_2)\cdot {d\over d\tau_2}q_{k_3}^*(\tau_2) \cdot q_{k_4}^*(\tau_2) \right] \nonumber \\
&&+ \int_{-\infty}^0 {d\tau_1\over \tau_1}  \cdot q_{k_{12}}(\tau_1)\cdot {d\over d\tau_1}  q_{k_1}(\tau_1) \cdot q_{k_2}(\tau_1) \nonumber \\
&&\times \left[ -2 ({\bf k}_{12}\cdot {\bf k}_2) ({\bf k}_3\cdot {\bf k}_4)  \int_{-\infty}^0 {d\tau_2\over \tau_2} \cdot {d\over d\tau_2} q_{k_{12}}^* (\tau_2)\cdot q_{k_3}^*(\tau_2)q_{k_4}^*(\tau_2)  \right. \nonumber \\
&&\left. \left. - 4 ({\bf k}_{12}\cdot {\bf k}_2) ({\bf k}_{12}\cdot {\bf k}_4)  \int_{-\infty}^0 {d\tau_2\over \tau_2} \cdot q_{k_{12}}^* (\tau_2)\cdot {d\over d\tau_2}q_{k_3}^*(\tau_2) \cdot q_{k_4}^*(\tau_2) \right] \right\} \nonumber \\
&&+23 \ \hbox{perms}.
\n
Eq.(\ref{hhq}) is
\m
&&-2 (2\pi)^3\delta^{(3)}(\sum_{i=1}^4{\bf k}_i) \cdot {\beta_1^2\over 4M^4H^2} \cdot q_{k_1}^*(0) q_{k_2}^*(0) q_{k_3}^*(0)q_{k_4}^*(0) \times \nonumber \\
&&\hbox{Re}\ \left\{ \int_{-\infty}^0 {d\tau_1\over \tau_1} {d\over d\tau_1} q_{k_{12}}^*(\tau_1)\cdot q_{k_3}(\tau_1)q_{k_4}(\tau_1) \right. \nonumber \\
&&\times \left[ ({\bf k}_1\cdot {\bf k}_2) ({\bf k}_3\cdot {\bf k}_4)  \int_{-\infty}^{\tau_1} {d\tau_2\over \tau_2} \cdot {d\over d\tau_2} q_{k_{12}} (\tau_2)\cdot q_{k_1}(\tau_2)q_{k_2}(\tau_2)  \right. \nonumber \\
&&\left. - 2 ({\bf k}_{12}\cdot {\bf k}_2) ({\bf k}_3\cdot {\bf k}_4)    \int_{-\infty}^{\tau_1} {d\tau_2\over \tau_2} \cdot q_{k_{12}} (\tau_2)\cdot {d\over d\tau_2}q_{k_1}(\tau_2) \cdot q_{k_2}(\tau_2) \right] \nonumber \\
&&+ \int_{-\infty}^0 {d\tau_1\over \tau_1}  \cdot q_{k_{12}}^*(\tau_1)\cdot {d\over d\tau_1}  q_{k_3}(\tau_1) \cdot q_{k_4}(\tau_1) \nonumber \\
&&\times \left[ 2 ({\bf k}_1\cdot {\bf k}_2) ({\bf k}_{12}\cdot {\bf k}_4)  \int_{-\infty}^{\tau_1} {d\tau_2\over \tau_2} \cdot {d\over d\tau_2} q_{k_{12}} (\tau_2)\cdot q_{k_1}(\tau_2)q_{k_2}(\tau_2)  \right. \nonumber \\
&&\left. \left. - 4 ({\bf k}_{12}\cdot {\bf k}_2) ({\bf k}_{12}\cdot {\bf k}_4)  \int_{-\infty}^{\tau_1} {d\tau_2\over \tau_2} \cdot q_{k_{12}} (\tau_2)\cdot {d\over d\tau_2}q_{k_1}(\tau_2) \cdot q_{k_2}(\tau_2) \right] \right\} \nonumber \\
&&+23 \ \hbox{perms}.
\n
We can do the similar calculations for the term with $\beta_2$ and the cross case.

For the scalar-exchange diagram, $T_\zeta(k_1,k_2,k_3,k_4)$ also depends on $k_{ij}$.  However, in general, we cannot effectively define the non-Gaussianity parameter $\tau_{NL}$ because the coefficients for different ${1\over k_{ij}^3k_j^3k_l^3}$ might be different for a given momenta configuration. Based on the symmetry of the four-point correlation function, we define three new parameters $\tau^{(i)}$ with $i=1,2,3$ for the general equilateral shape trispectrum ($k_1=k_2=k_3=k_4\equiv k$),
\m
\langle \zeta_{{\bf k}_1} \zeta_{{\bf k}_2} \zeta_{{\bf k}_3} \zeta_{{\bf k}_4} \rangle  \supset  (2\pi)^9 \delta^{(3)}(\sum_{i=1}^4{\bf k}_i) \cdot {1\over 16} P_\zeta^3 \times \Xi(k_i,k_{ij}),
\label{ttt}
\n
where
\m
\Xi(k_i,k_{ij})&=& \tau^{(1)} \({1 \over k_{12}^3 k_2^3 k_3^3}+{1 \over k_{12}^3 k_2^3 k_4^3}+{1 \over k_{21}^3 k_1^3 k_3^3}+{1 \over k_{21}^3 k_1^3 k_4^3}+(1\leftrightarrow 3,2\leftrightarrow 4)\)  \nonumber \\
&+&\tau^{(2)}\({1 \over k_{13}^3 k_3^3 k_2^3}+{1 \over k_{13}^3 k_3^3 k_4^3}+{1 \over k_{31}^3 k_1^3 k_2^3}+{1 \over k_{31}^3 k_1^3 k_4^3}+(1\leftrightarrow 2,3\leftrightarrow 4)\) \nonumber\\
&+&\tau^{(3)} \({1 \over k_{14}^3 k_4^3 k_2^3}+{1 \over k_{14}^3 k_4^3 k_3^3}+{1 \over k_{41}^3 k_1^3 k_2^3}+{1 \over k_{41}^3 k_1^3 k_3^3}+(1\leftrightarrow 2,3\leftrightarrow 4)\) . \nonumber
\n
For the special equilateral shape ``ST" in which $\theta_{12}=\theta_{13}=\theta_{14}=\cos^{-1}(-1/3)$, $k_{12}=k_{13}=k_{14}={2\over \sqrt{3}}k$, $\tau^{(1)}=\tau^{(2)}=\tau^{(3)}$ and then we can we can effectively define $\tau_{NL}^{ST}$ as follows
\m
\langle \zeta_{{\bf k}_1} \zeta_{{\bf k}_2} \zeta_{{\bf k}_3} \zeta_{{\bf k}_4} \rangle \supset (2\pi)^9 \delta^{(3)}(\sum_{i=1}^4{\bf k}_i) \cdot {1\over 16} \tau_{NL}^{ST} P_\zeta^3 \cdot \({1 \over k_{12}^3 k_2^3 k_3^3}+23 \ \hbox{perms}.\),
\n
where $\tau_{NL}^{ST}\equiv \tau^{(1)}=\tau^{(2)}=\tau^{(3)}$ at ``ST" shape.

\begin{figure}[h]
\begin{center}
\includegraphics[width=12cm]{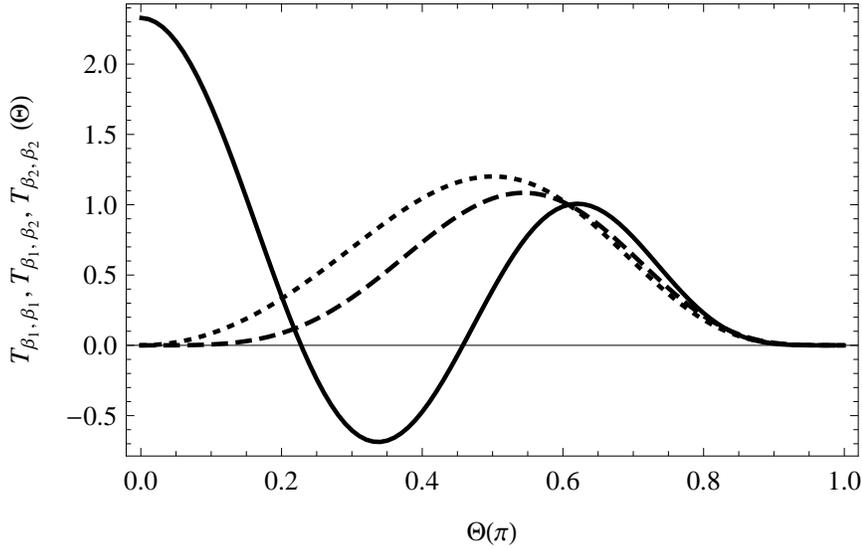}
\end{center}
\caption{The functions $T_{\beta_1,\beta_1}(\Theta)$, $T_{\beta_1,\beta_2}(\Theta)$ and $T_{\beta_2,\beta_2}(\Theta)$ which correspond to the solid, dotted and dashed lines. }
\label{taunlb}
\end{figure}

Considering $\zeta_{\bf k}=-{H\over M^2}Q_{\bf k}$,
and
\e
{d\over d\tau}q_k(\tau)=-\sqrt{\pi\over 8} {\alpha H^2\over M} k^2 (-\tau)^{5/2} H_{-1/4}^{(1)} \({\alpha H k^2\over 2M}\tau^2\),
\q
we find
\m
\tau^{(i)}&\simeq& 0.0437 \cdot \beta_1^2\cdot {M^2\over H^2}\cdot \alpha^{-2} \cdot T_{\beta_1,\beta_1}(\Theta_i)+0.492  \cdot \beta_2^2\cdot {M^2\over H^2}\cdot \alpha^{-4} \cdot T_{\beta_2,\beta_3}(\Theta_i) \nonumber \\
&-&0.156 \cdot  \beta_1  \beta_2 \cdot {M^2\over H^2}\cdot \alpha^{-3} \cdot T_{\beta_1,\beta_2}(\Theta_i),
\n
where $i=1,2,3$, and $\Theta_1=\theta_{12}$, $\Theta_2=\theta_{13}$, $\Theta_3=\theta_{14}$.
Combining with WMAP normalization, we obtain
\m
\tau^{(i)}&\simeq&1.75\times 10^4\cdot \beta_1^2\cdot \alpha^{-16/5}\cdot T_{\beta_1,\beta_1}(\Theta_i)+1.97\times 10^5\cdot \beta_2^2\cdot \alpha^{-26/5}\cdot T_{\beta_2,\beta_2}(\Theta_i) \nonumber \\
&-&6.25\times 10^4\cdot \beta_1\beta_2 \cdot \alpha^{-21/5}\cdot T_{\beta_1,\beta_2}(\Theta_i),
\n
where  $T_{\beta_1,\beta_1}(\Theta)$, $T_{\beta_1,\beta_2}(\Theta)$ and $T_{\beta_2,\beta_2}(\Theta)$ are showed in Fig. \ref{taunlb}.
Here we normalized these three functions at $\Theta=\cos^{-1}(-1/3)$. From Fig. \ref{taunlb}, we see that the contribution by $\dot Q (\nabla Q)^2$ is quite different from that by $\nabla^2 Q (\nabla Q)^2$.
Roughly speaking, $\tau^{(i)}$ for $i=1,2,3$ are peaked around the special tetrahedron shape where $\theta_{12}=\theta_{13}=\theta_{14}=\cos^{-1}(-1/3)$. For the local shape trispectrum, $T_\zeta(k_1,k_2,k_3,k_4)$ blows up for the shape of $k_{ij}\rightarrow 0$ which corresponds to $\theta_{ij}\rightarrow \pi$. But here $\tau^{(i)}$ goes to zero when $\theta_{ij}\rightarrow \pi$. Therefore the scalar-exchange diagram in the ghost inflation cannot generate a local shape trispectrum.
For the ``ST" shape, we obtain
\m
\tau_{NL}^{ST}&\simeq&1.75\times 10^4\cdot \beta_1^2\cdot \alpha^{-16/5}+1.97\times 10^5\cdot \beta_2^2\cdot \alpha^{-26/5} \nonumber \\
&-&6.25\times 10^4\cdot \beta_1\beta_2 \cdot \alpha^{-21/5}
\n
which is positive definite.

Actually there is not a distinguishing feature to define the analogous parameters to $g_{NL}^{loc.}$ and $\tau_{NL}^{loc.}$ for the non-local shape trispectrum.
Since the trispectrum from the scalar-exchange diagram in ghost inflation does not blow up like that for the local shape trispectrum with non-zero $\tau_{NL}^{loc.}$ at the limit of $k_{ij}\rightarrow 0$, it is hard to make the meaning of $\tau_{NL}^{ST}$ clear. For example, we can also calculate the effective $g_{NL}^{equil.}$ defined in Eq. (\ref{defgnl}) for the trispectrum from scalar-exchange diagram.
Comparing Eq. (\ref{ttt}) with (\ref{defgnl}), we obtain
\m
g_{NL}^{equil.}={25\over 108}\Xi(k_i,k_{ij})\times {\prod_{i=1}^4 k_i^3\over \sum_{i=1}^4 k_i^3}= {25\over 432} k^9 \Xi(k_i,k_{ij}),
\n
here we focus on the equilateral shape with $k_1=k_2=k_3=k_4\equiv k$. For the ``ST" shape, $k_{12}=k_{13}=k_{14}=2k/\sqrt{3}$ and then
\e
\Xi^{ST}={9\sqrt{3}\over k^9}\tau_{NL}^{ST}.
\q
Therefore
\e
g_{NL}^{ST}={25\sqrt{3}\over 48}\tau_{NL}^{ST},
\label{cgt}
\q
where $g_{NL}^{ST}$ is $g_{NL}^{equil.}$ evaluated at ``ST" shape.
Similarly, we can also define an effective $\tau_{NL}^{ST}$ for the contact diagram as
\e
\tau_{NL}^{ST}=-9.61\times 10^3\cdot \gamma\cdot \alpha^{-16/5}.
\q
Here $\gamma$ denotes that it is contributed from the contact diagram.

\begin{figure}[h]
\begin{center}
\includegraphics[width=10cm]{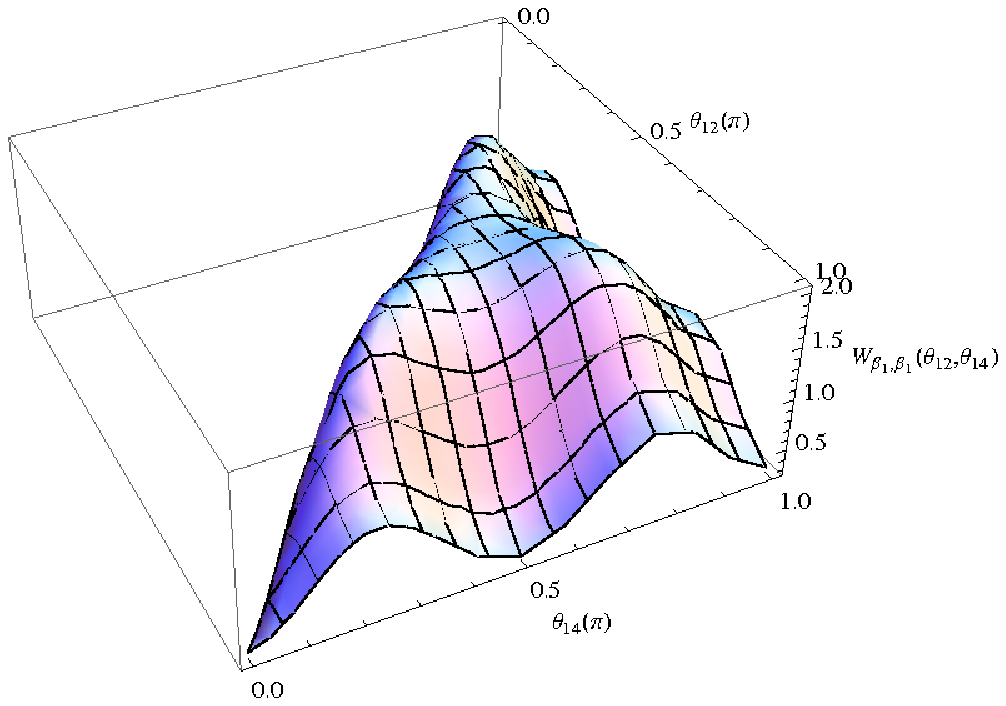}\\
\includegraphics[width=10cm]{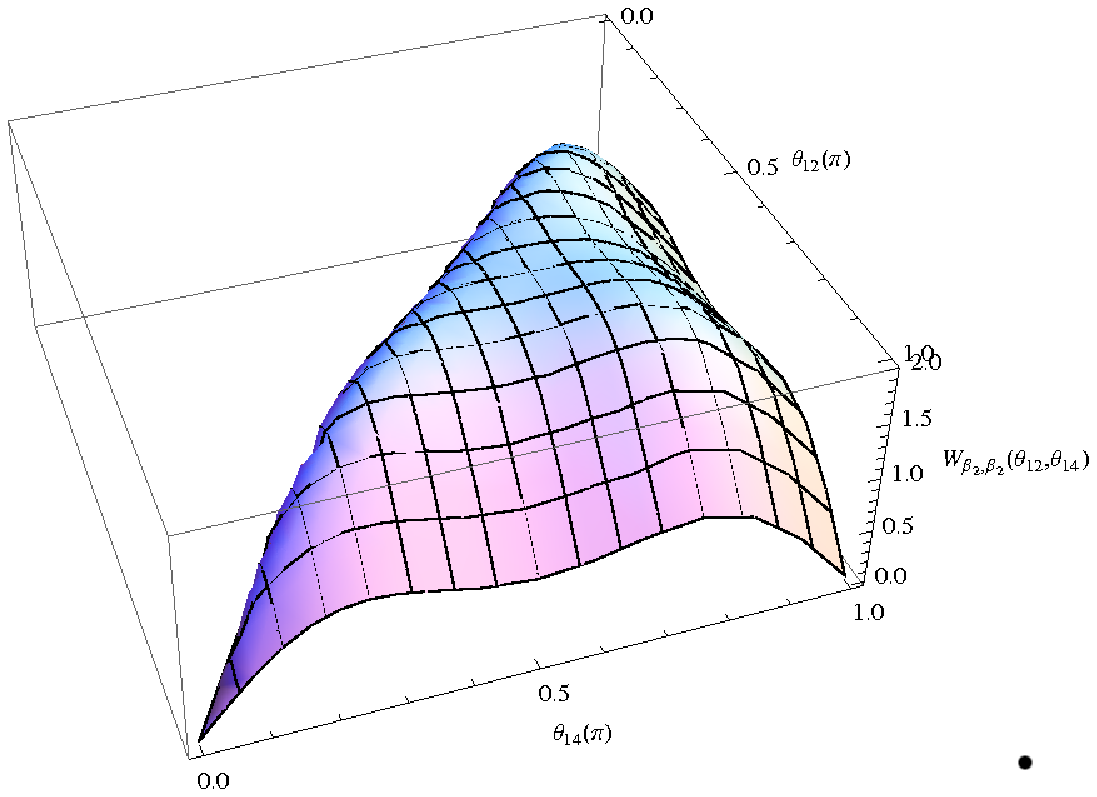}\\
\includegraphics[width=10cm]{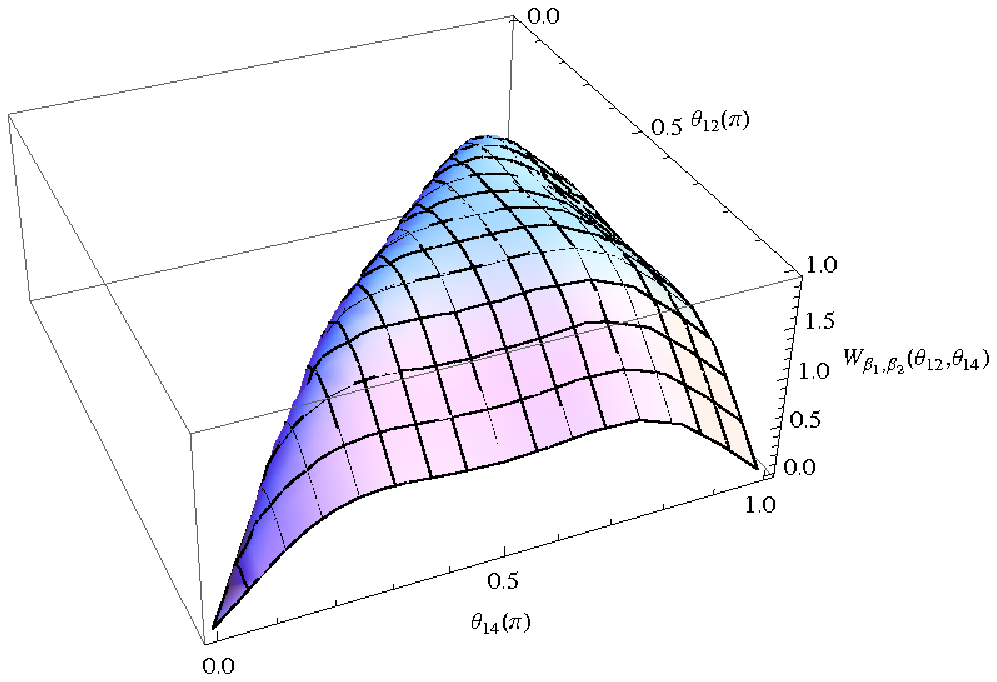}
\end{center}
\caption{The functions $W_{\beta_1,\beta_1}(\theta_{12},\theta_{14})$, $W_{\beta_2,\beta_2}(\theta_{12},\theta_{14})$ and $W_{\beta_1,\beta_2}(\theta_{12},\theta_{14})$. }
\label{www}
\end{figure}

For general equilateral shape, we have
\m
{k^2_{12}\over k^2}&=&2(1+\cos\theta_{12}),\\
{k^2_{13}\over k^2}&=&2(1+\cos\theta_{13}),\\
{k^2_{14}\over k^2}&=&2(1+\cos\theta_{14}),
\n
and thus
the effective non-gaussianity parameter contributed from the scalar-exchange diagram in ghost inflation is
\m
g_{NL}^{equil.}&=& {25\over 54} \[{\tau^{(1)}\over (2(1+\cos\theta_{12}))^{3/2}}+{\tau^{(2)}\over (2(1+\cos\theta_{13}))^{3/2}}+{\tau^{(3)}\over (2(1+\cos\theta_{14}))^{3/2}}\] \nonumber \\
&\simeq&8.1\times 10^3\cdot \beta_1^2\cdot \alpha^{-16/5}\cdot W_{\beta_1,\beta_1}(\theta_{12},\theta_{14})+9.1\times 10^4\cdot \beta_2^2\cdot \alpha^{-26/5}\cdot W_{\beta_2,\beta_2}(\theta_{12},\theta_{14}) \nonumber \\
&-&2.9\times 10^4\cdot \beta_1\beta_2 \cdot \alpha^{-21/5}\cdot W_{\beta_1,\beta_2}(\theta_{12},\theta_{14}),
\n
where
\m
W_{\beta_1,\beta_1}(\theta_{12},\theta_{14})&=&{T_{\beta_1,\beta_1}(\theta_{12})\over (2(1+\cos\theta_{12}))^{3/2}}+{T_{\beta_1,\beta_1}(\theta_{13})\over (2(1+\cos\theta_{13}))^{3/2}}+{T_{\beta_1,\beta_1}(\theta_{14})\over (2(1+\cos\theta_{14}))^{3/2}},\nonumber \\
W_{\beta_1,\beta_2}(\theta_{12},\theta_{14})&=&{T_{\beta_1,\beta_2}(\theta_{12})\over (2(1+\cos\theta_{12}))^{3/2}}+{T_{\beta_1,\beta_2}(\theta_{13})\over (2(1+\cos\theta_{13}))^{3/2}}+{T_{\beta_1,\beta_2}(\theta_{14})\over (2(1+\cos\theta_{14}))^{3/2}},\nonumber \\
W_{\beta_2,\beta_2}(\theta_{12},\theta_{14})&=&{T_{\beta_2,\beta_2}(\theta_{12})\over (2(1+\cos\theta_{12}))^{3/2}}+{T_{\beta_2,\beta_2}(\theta_{13})\over (2(1+\cos\theta_{13}))^{3/2}}+{T_{\beta_2,\beta_2}(\theta_{14})\over (2(1+\cos\theta_{14}))^{3/2}},\nonumber
\n
which are shown in Fig. \ref{www} respectively.
The parameter $g_{NL}^{equil.}$ is shape-dependent and $g_{NL}^{equil.}$ goes to zero when $\theta_{12}$ and/or $\theta_{14}$ go to $\pi$.

\section{Discussions}

In this paper we calculate the trispectrum from the contact
and scalar-exchange diagram in ghost inflation. The shape
of trispectrum generated by the contact diagram is quite
different from that produced by the scalar-exchange
diagram. Roughly speaking, the former is peaked at the
shapes L1, L2, and L3 and minimized at the shape ST, but
the later is the reverse.

In this paper, we introduce a new shape, so-called ``planar
mirror symmetric quadrangle" shape. This shape is very
useful to distinguish the term with $\tau_{NL}^{loc.}$ in
trispectrum from that with $g_{NL}^{loc.}$: in the case of
$\theta_2=(\pi-\theta_1)/2$ and $k_1=k_2=k_3=k_4$, only the
term with $\tau_{NL}^{loc.}$ blows up in the limit of
$\theta_1\rightarrow 0$; but both of them blow up in the
limit of $\theta_1\rightarrow 0$ when $\theta_2=\pi/3$.

For the local form non-Gaussianity, the trispectrum is
completely characterized by two independent parameters
$\tau_{NL}^{loc.}$ and $g_{NL}^{loc.}$. However, for the
general single-field inflation, including ghost inflation,
the trispectrum does not blow up in the limit of
$k_i\rightarrow 0$ and/or $k_{ij}\rightarrow 0$. For the
non-local form trispectrum, the analogous to
$\tau_{NL}^{loc.}$ can be defined only for the shape ``ST",
but there is not a definite feature to distinguish
$\tau_{NL}^{ST}$ from $g_{NL}^{equil.}$. In Sec. 4.2, we
clearly illustrate that $\tau_{NL}^{ST}$ is nothing but
$g_{NL}^{ST}$ which is $g_{NL}^{equil.}$ evaluated at ``ST"
shape.

The trispectrum in the ghost inflation is quite different
from that in the single-field inflation without breaking
Lorentz symmetry. For example, DBI inflation, a typical
inflation in string theory, predicts a definite positive
value of $\tau_{NL}^{ST}$, but it can be either positive or
negative in ghost inflation. The trispectrum potentially
provides a very useful discriminator for inflation models.
However, how to define some observables which can be
directly used to compare to the observational data is still
an open question for the non-local shape trispectrum. Here
we need to emphasis that the local form non-Gaussianity is
much more sensitive to the cosmological observations
\cite{Kogo:2006kh,Jeong:2009vd,Desjacques:2009jb,Chingangbam:2009vi,Vielva:2009jz,Maggiore:2009hp}.
For example, the uncertainties of $f_{NL}^{loc.}$,
$\tau_{NL}^{loc.}$ and $g_{NL}^{loc.}$ will be reduced to
$\Delta f_{NL}^{loc.}=5$, $\Delta \tau_{NL}^{loc.}\simeq
560$ and $\Delta g_{NL}^{loc.}\simeq 1.3\times 10^4$ by
Planck. A convincing detection of non-Gaussianity with
local or non-local form will have a profound implication
for the physics in the early universe.

\vspace{1.5cm}

Note added: After we submitted our paper to arXiv, a paper
\cite{Izumi:2010wm} working on the same topic appeared in
arXiv as well. Actually our discussions are more complete.
We considered contributions to the trispectrum from the
terms of $\dot Q(\nabla Q)^2$ and $\nabla^2Q (\nabla Q)^2$
and cross case in the scalar-exchange diagram. However in
\cite{Izumi:2010wm} the authors only took $\dot Q(\nabla
Q)^2$ into account.

\vspace{1.5cm}

\noindent {\bf Acknowledgments}

\vspace{.5cm}

We would like to thank P.~Chingangbam and P.~Yi for useful discussions. This work is supported by the project of Knowledge Innovation Program of Chinese Academy of Science.





\end{document}